\documentclass{emulateapj}

\usepackage{graphicx}
\usepackage{datetime}

\slugcomment{Received 2014 January 6; accepted 2014 March 10}

\shorttitle{Asymmetric fundamental band CO lines}
\shortauthors{Reg\'aly et al.}

\begin{document}

\title{Asymmetric fundamental band CO lines as a sign of an embedded giant planet}

\author{Zs. Reg\'aly$^{1,2}$, S. Kir\'aly$^1$, and L. L. Kiss$^{1,3}$}
\affil{$^1$Konkoly Observatory, Research Center for Astronomy and Earth Sciences, P.O. Box 67, H-1525 Budapest, \\Hungary; regaly@konkoly.hu}
\affil{$^2$ELTE Gothard-Lend\"ulet Research Group, H-9704 Szombathely, Szent Imre Herceg u. 112, Hungary}
\affil{$^3$Sydney Institute for Astronomy, School of Physics A28, University of Sydney, NSW 2006, Australia}

\begin{abstract}
We investigate the formation of double-peaked asymmetric line profiles of CO in the fundamental band  spectra emitted by young ($1\textrm{--}5$\,Myr) protoplanetary disks hosted by a $0.5\textrm{--}2\,M_\odot$ star. Distortions of the  line profiles can be caused by the gravitational perturbation of an embedded giant planet with $q=4.7\times10^{-3}$ stellar-to-planet mass ratio. Locally isothermal, two-dimensional hydrodynamic simulations show that the disk becomes globally eccentric inside the planetary orbit with stationary $\sim0.2\textrm{--}0.25$ average eccentricity after $\sim2000$ orbital periods. For orbital distances $1\textrm{--}10$\,AU, the disk eccentricity is peaked inside the region where the fundamental band of CO is thermally excited. Hence, these lines become sensitive indicators of the embedded planet via their asymmetries (both in flux and wavelength). We find that the line shape distortions (e.g., distance, central dip, asymmetry, and positions of peaks) of a given transition depend on the excitation energy (i.e., on the rotational quantum number $J$). The magnitude of line asymmetry is increasing/decreasing with $J$ if the planet orbits inside/outside the CO excitation zone ($R_\mathrm{CO}\leq3,\,5$ and $7\,\rm AU$ for a  $0.5,\,1,$ and $2\,M_\odot$ star, respectively), thus one can constrain the orbital distance of a giant  planet by determining  the slope of the peak asymmetry--$J$ profile. We conclude that the presented spectroscopic phenomenon can be used to test the predictions of planet formation theories by pushing the age limits for detecting the youngest planetary systems.
\end{abstract}

\keywords{accretion, accretion disks --- hydrodynamics --- line: formation --- methods: numerical --- protoplanetary disks --- techniques: spectroscopic}

\section{Introduction}

Protoplanetary disks are expected to emit symmetric double-peaked molecular lines in  infrared wavelengths as a result of the Keplerian angular velocity distribution of gas parcels \citep{HorneMarsh1986}. Contrary to this, asymmetric line  profiles have been observed in the fundamental band of CO for several young ($1\textrm{--}5$\,Myr) protoplanetary disks \citep{Pontoppidanetal2008,BlakeBoogert2004,Dentetal2005,Salyketal2009,Brownetal2013}.

According to the grid-based numerical simulations of \cite{KleyDirksen2006}, disk eccentricity can be excited locally, near the gap opened by an embedded giant planet. As the orbital velocity of the gas parcels is highly supersonic in accretion disks, gas parcels in eccentric orbits result in supersonic velocity deviations in comparison to the circular Keplerian case. \cite{Horne1995} has shown that supersonic turbulence can cause observable line profile distortions in the molecular spectra of protoplanetary disks. \citet{Regalyetal2010} have demonstrated that the double-peaked Keplerian  line profiles  in the fundamental  band of CO are distorted  due to the excitation of disk eccentricity in the vicinity of the gap carved by a close orbiting ($\leq1\,\rm AU$)  giant planet ($M_\mathrm{p}>1\,M_{Jup}$), which allows an indirect detection of the planet.

The theory of resonant excitation mechanisms in accretion disks of \cite{Lubow1991} predicts that the circumstellar disks of close-separation young binaries become fully eccentric due to the orbiting companion.  Locally isothermal numerical simulations have  confirmed this  
\citep{Kleyetal2008,KleyNelson2008,Paardekooperetal2008,Marzarietal2009,Regalyetal2011}. However, only fast cooling and low-mass disks favor the excitation of disk eccentricity with radiative and self-gravitating disk approximations \citep{Marzarietal2012,MullerKley2012}. Radiative three-dimensional SPH  simulations of \citet{PicognaMarzari2013} have also revealed $\sim0.1$ eccentricity excitation of circumstellar disks in binaries. Nonetheless, a disk having a non-constant radial eccentricity profile with $\sim0.2\textrm{--}0.3$ average eccentricity emits clearly asymmetric fundamental band CO lines \citep{Regalyetal2011}. 

Since no close stellar mass companions were found for the majority of disks that emits asymmetric lines, it is worth investigating whether a giant planet is able to excite global disk eccentricity inside its orbit where the CO is thermally excited. We present the excitation of  global disk eccentricity by means of two-dimensional hydrodynamical simulations, which leads to the formation of asymmetric double-peaked line profiles of CO.

\section{Formation of an Eccentric Disk}

\subsection{Hydrodynamical Simulations}

The interaction between the disk and the embedded giant planet with stellar-to-planet mass ratio of $q=4.7\times10^{-3}$ (corresponding to $2.5,\,5$ and $10\,M_\mathrm{Jup}$ for a $0.5,1$ and $2\,M_{\odot}$ central star, respectively) orbiting at fixed distance $a_\mathrm{p}=1,\,3,\,5,\,7,$ and $10$\,AU has been investigated by two-dimensional grid-based, global hydrodynamic simulations. We used the GPU supported version of the code {\small FARGO} \citep{Masset2000}, which solves the vertically integrated continuity and Navier--Stokes equations numerically, using locally isothermal approximation, for which case the disk's radial temperature profile is $T(R)\sim R^{-1}$.

We adopt as the unit of length $1\,\rm AU$ and the unit of mass is the solar mass $1\,M_\odot$. The gravitational constant is set to unity, thus the unit of time equals the orbital period of a planet at $1\,\rm AU$, which is $2\pi$. We use a frame that corotates with the planet.

The disk's viscosity is approximated by the $\alpha$ prescription \citep{ShakuraSunyaev1973} with a canonical value of $\alpha=0.001$. The gas density has a power-law profile $\Sigma(R)=\Sigma_0R^{-0.5}$ initially, such that $0.01\,M_\odot$ mass is confined within 30\,AU. We assume a flat disk model, such that the local scale height of the disk is $H(R) = hR$, where $h=0.05$.

At the inner and outer boundaries, the damping boundary condition \citep{deValBorroetal2006} is applied. The spatial extension of the computational domain is $0.2\textrm{--}15$\,AU\footnote{In order to excite the global disk eccentricity for the $a_\mathrm{p}=1$\,AU model, the inner boundary of the disk must be set to $\leq0.05$\,AU.}, consisting of $N_R=256$ logarithmically distributed radial and $N_\phi=512$ equidistant azimuthal grid cells.

To take the disk thickness into account, we use $\epsilon H(a_\mathrm{p})$ as the smoothing of the gravitational potential of the planet with $\epsilon=0.6$ \citep{Kleyetal2012}.

\subsection{Excitation of the Disk Eccentricity}

\begin{figure}
	\centering
	\includegraphics[width=8.5cm]{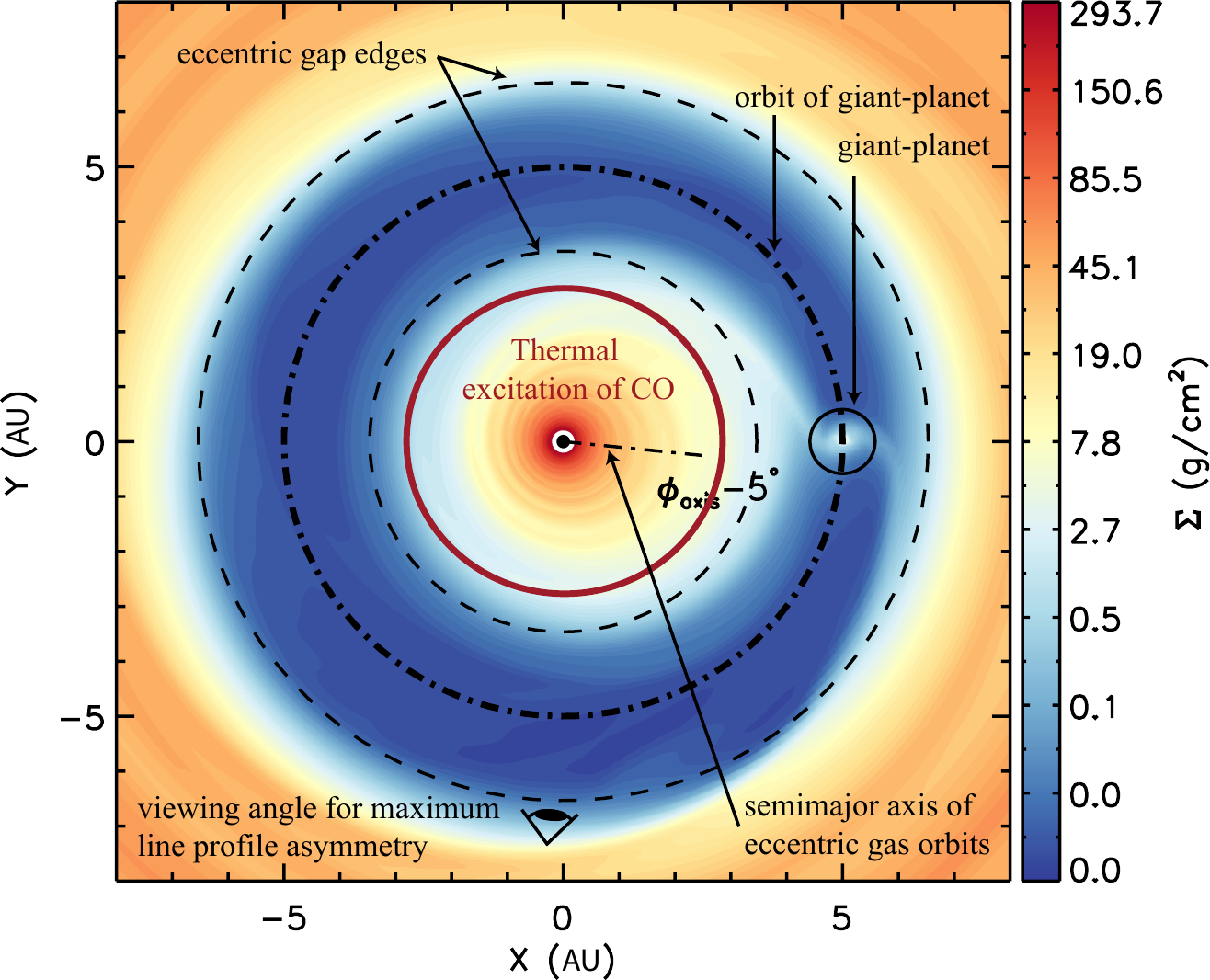}
	\caption{Disk surface density distribution in model $a_\mathrm{p}=5\, \rm AU$ at $t=40\times10^3$\,yr, when a quasi-static eccentric disk state has already been developed.}
	\label{fig:eccdisk}
\end{figure}

Thanks to the high computational performance provided by the GPUs, we were able to follow the viscous evolution to $t\simeq40\times10^3$\,yr (corresponding to several thousand orbits of the embedded planet), much longer than in our previous study \citep{Regalyetal2010}. The giant planet opens a gap in a couple of orbits, then significant disk eccentricity develops at the gap outer edge in $\sim500$ orbits.  Integrating further, we found that the disk also becomes eccentric inside the planetary orbit (Figure\,\ref{fig:eccdisk}).

\begin{figure}
	\centering
	\includegraphics[width=\columnwidth]{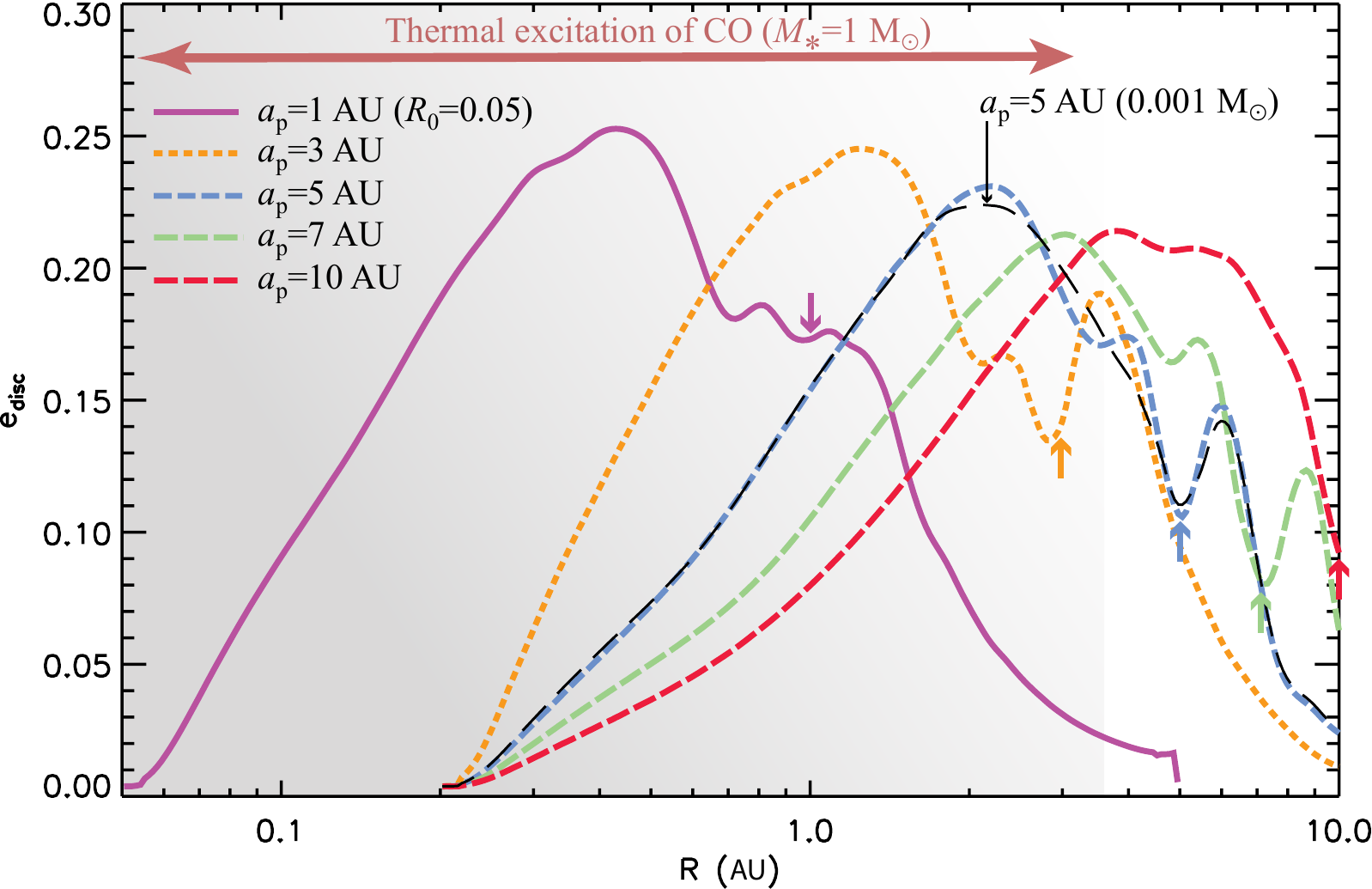}
	\caption{Disk eccentricity profiles in models $1\,\mathrm{AU} \leq a_\mathrm{p}\leq 10\,\mathrm{AU}$. An additional simulation  is also shown, where $M_\mathrm{disk}=0.001\,M_\odot$ for model $a_\mathrm{p}=5\,\rm AU$. Planetary orbital distances indicated by the arrows.}
	\label{fig:decprof}
\end{figure}

The radially averaged disk eccentricity ($\langle {e}_\mathrm{disk}\rangle$) evolves in time, first reaching a maximum value of $\sim0.3$, then declining, and a quasi-static state develops with $\langle e_\mathrm{disk}\rangle\simeq0.2$ within $\sim2000$ orbits\footnote{The evolution of $\langle {e}_\mathrm{disk}\rangle$ is found to be  very similar to that of models simulating the perturbation of a stellar companion on the circumprimary disk in a close binary system \citep{Regalyetal2011}.}.  Figure\,\ref{fig:decprof} shows the radial profiles of disk eccentricity in the quasi-static eccentric state.  Note also that the disk eccentricity peaks at $0.2\textrm{--}0.25$ inside the planetary orbit at $R_\mathrm{e-max}/a_\mathrm{p}\simeq0.4$  for all models. Although our disk is relatively  massive, the disk eccentricity profiles are found to be very similar assuming significantly lower $0.001\,M_\odot$ disk mass (shown in  Figure\,\ref{fig:decprof}  for model $a_\mathrm{p}=5\,\rm AU$).

\section{Origin of Asymmetric CO Lines}

\begin{figure*}
	\centering
	\includegraphics[width=\columnwidth]{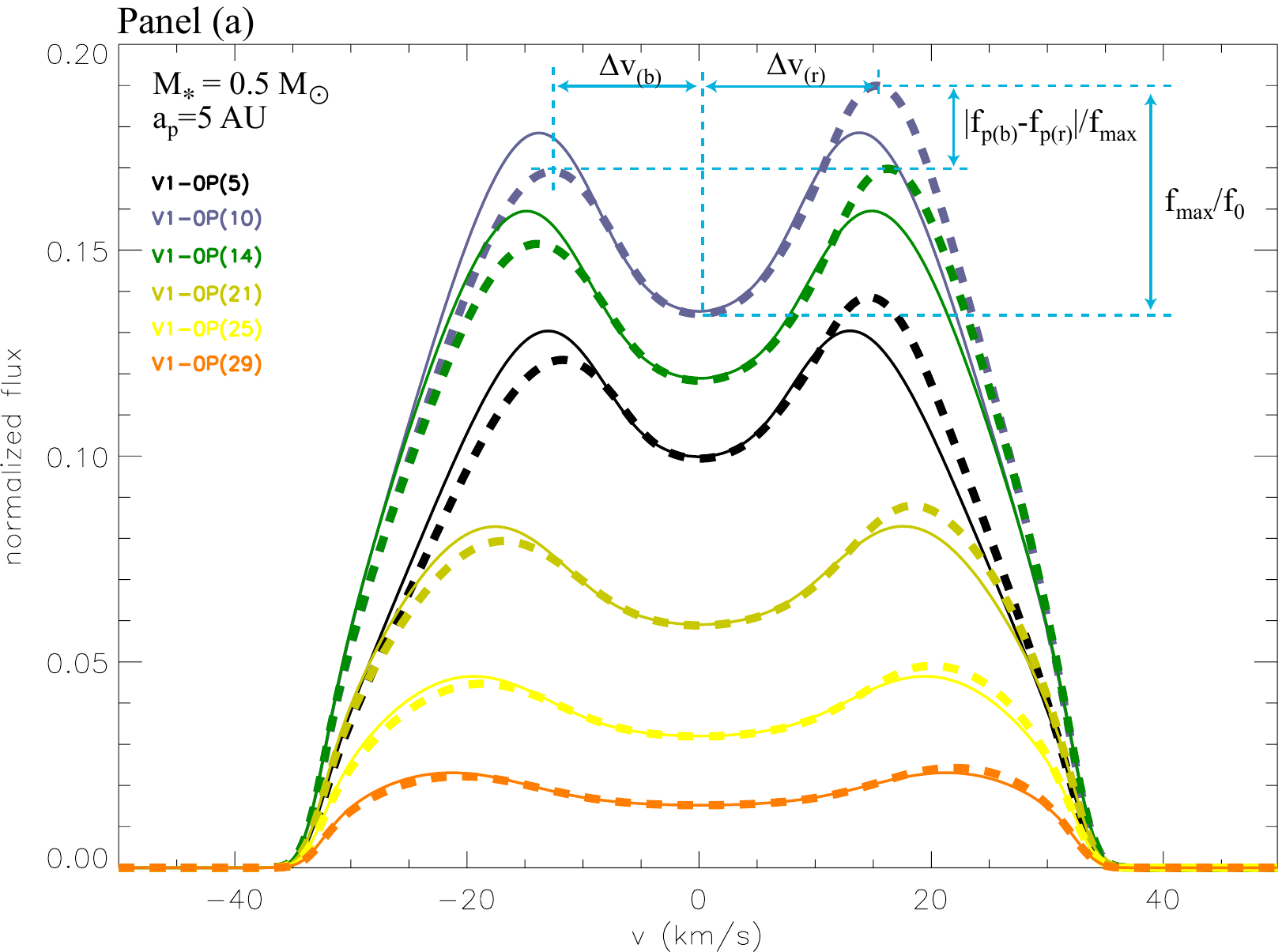}
	\includegraphics[width=\columnwidth]{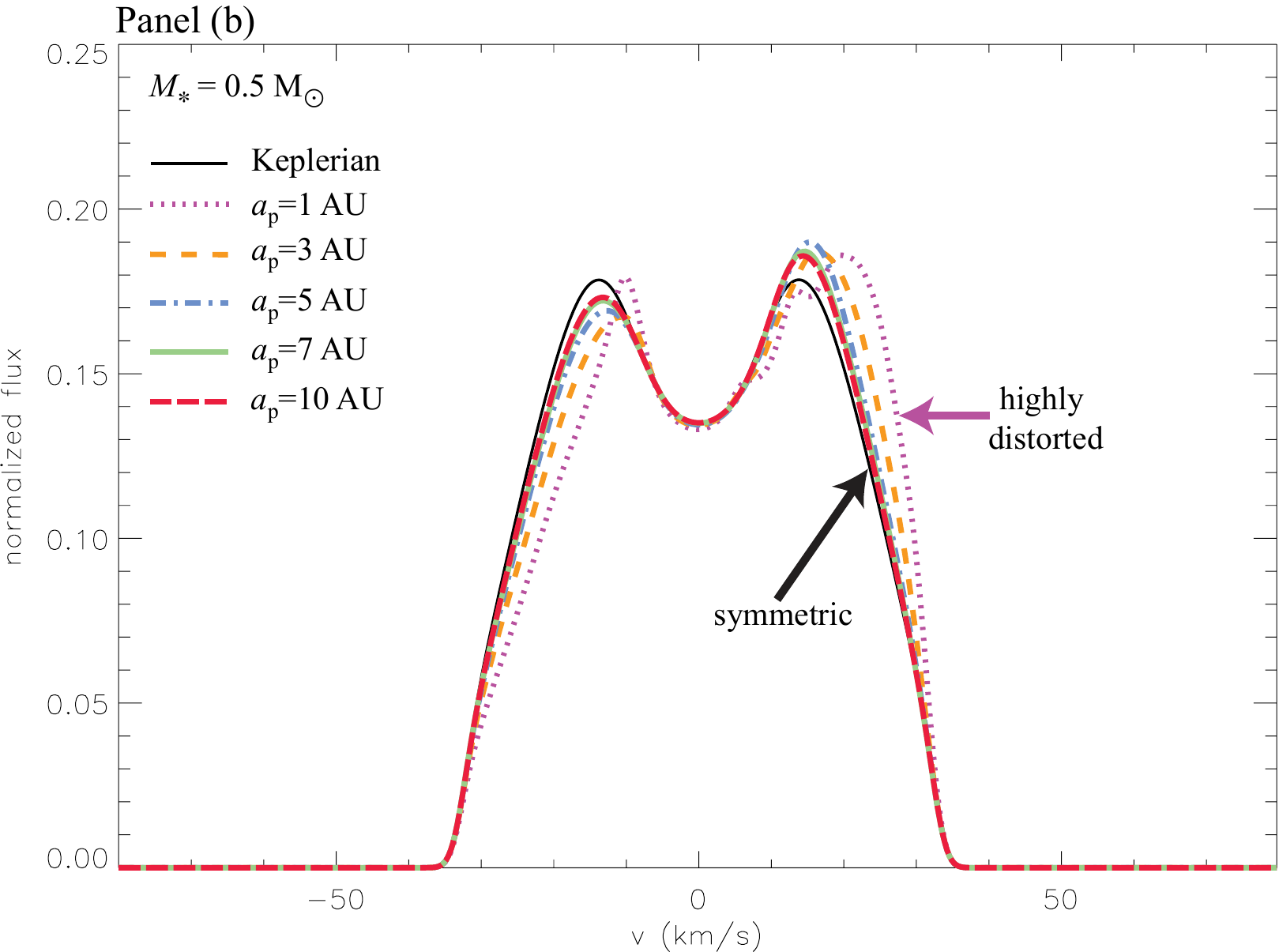}
	\includegraphics[width=\columnwidth]{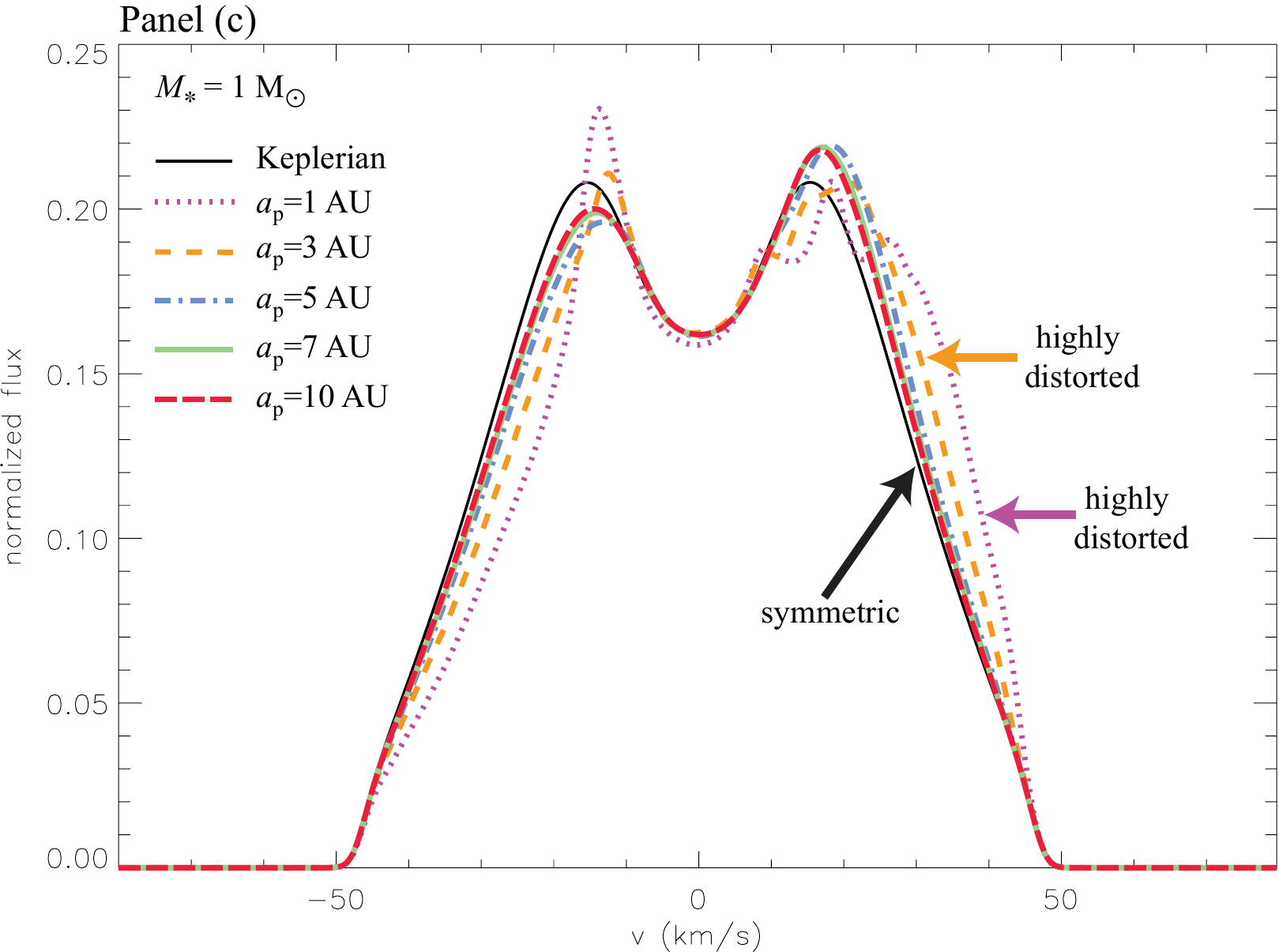}
	\includegraphics[width=\columnwidth]{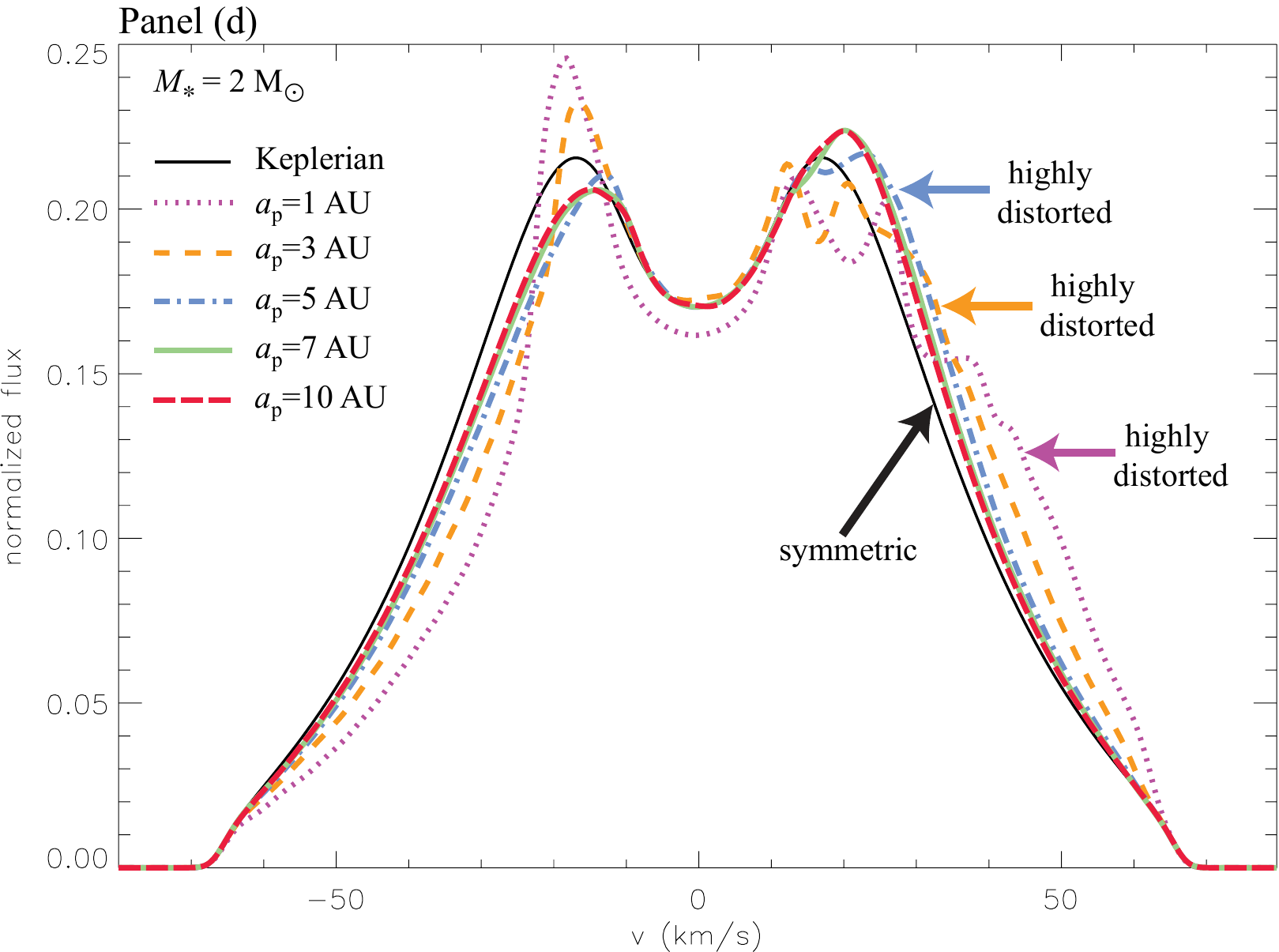}
	\caption{Continuum-normalized (and vertically shifted by 1) line profiles in the fundamental band of the CO molecule. Panel (a): selected non-blended {\it P}-branch transitions in model $a_\mathrm{p}=5\,\rm AU$ and $M_*=0.5\,M_\odot$. Panels (b)--(d): {\it P}(10) line profiles in models where $1\,\mathrm{AU}\leq a_\mathrm{p}\leq10\,\mathrm{AU}$ for $M_*=0.5,\,1$ and $2\,M_\odot$, respectively. Symmetric line profiles for an unperturbed "Keplerian" disk are also shown for comparison.}
	\label{fig:lineprofiles}
\end{figure*}

\subsection{Synthetic Spectral Model of CO Emission }

To calculate the CO lines, we used a previously developed semianalytical thermal excitation model \citep{Regalyetal2010}. For clarity, we summarize the main properties of that model. Our thermodynamical model is based on the double-layer disk model of \citet{ChiangGoldreich1997}. The disk is approximated by an optically thin hot atmosphere heated by the absorption of stellar irradiation by the dust and an optically thick interior heated by the atmosphere and accretion processes. 

The total emission is the sum of the stellar emission, atmospheric gas lines, dust continuum from the disk atmosphere, and interior. The stellar emission is calculated as a black-body radiation at the stellar surface temperature of $T_*=3800,\,4300$ and $4700$\,K based on the evolution tracks of a $0.5,\,1,$ and $2\,M_\odot$ at the age of $\sim1\textrm{--}5$\,Myr stars \citep{Siessetal2000}. The accretion rate is assumed to be $10^{-8}\,M_\odot yr^{-1}$ \citep{Fangetal2009}.

It is assumed that the dust consists of pure silicates with $1\,\mathrm{\mu m}$ grain size, for which case the mass absorption coefficient at visual and $\lambda=4.7\,\mathrm{\mu m}$ wavelengths are $\kappa_\mathrm{dust}=1780$ and $410\,\mathrm{cm^2\,g^{-1}}$, respectively \citep{DraineLee1984}. Due to settlement and photo-evaporation, larger grain sizes are not expected to be present in the disk atmosphere assuming no turbulent mixing-up from the disk interior. Spatially constant dust-to-gas and CO-to-gas mass ratios are used with $X_\mathrm{d}=10^{-2}$ and $X_\mathrm{CO}=4\times10^{-4}$, respectively.

The gas temperature is regulated by the collisions of molecules with the dust grains. We adopt perfect thermal coupling, as the number density of the gas in the tenuous disk atmosphere is $n_\mathrm{atm}\simeq10^{20}\,\mathrm{cm^{-3}}$ at 1\,AU, while thermal uncoupling occurs below $n_\mathrm{cr}\simeq 10^{14}\,\mathrm{cm^{-3}}$ \citep{ChiangGoldreich1997}. Note that for a completely opened gap, we found $n_\mathrm{atm}\simeq10^{16}\,\mathrm{cm^{-3}}$.

 The viewing angle (see eyepiece in Figure\,\ref{fig:decprof}) is set perpendicular to the major axis of the eccentric disk to obtain the strongest eccentricity signal\footnote{Changing the viewing angle by $\pm 90^\circ$ would result in no eccentricity signal.}. The eccentricity signal 
 is sensitive to the disk inclination angle ($i$) too.  For face-on ($i=0^\circ$), zero, or edge-on disk ($i=90^\circ$), the maximum\footnote{Without taking into account the disk self-obscuration, which would needlessly complicate our spectral model.} eccentricity signal would be observed; we assume an intermediate disk inclination, $i=45^\circ$.
 
The disk is depleted to $\Sigma_\mathrm{gap,d}\simeq3\times10^{-4}\,\mathrm{g\,cm^{-2}}$ in the dust inside the gap; however, our thermal model remains applicable there. Assuming a flat disk model, the grazing angle $\delta_\mathrm{1\,AU}\simeq1/500$, thus the optical depth along the incident stellar irradiation ($\tau_\mathrm{V}=\Sigma_\mathrm{gap,d}\kappa_\mathrm{dust,V}/\delta_\mathrm{1\,AU}$) is above $\sim3$ at optical wavelengths. Moreover, although the disk in the gap does not emit near-IR continuum while it does in our model, the continuum excess is negligible in the total disk flux. In a more realistic heating model, however, the temperature might be lower at  the inner edge of the gap due to its self-shadowing and higher at the outer edge due to the extensive illumination, which can cause additional disturbances in the line profiles \citep{Jang-CondellTurner2012}.

The two-dimensional computational domain of the CO spectral model is the same that of the hydrodynamical simulations. The optical depth at a given wavelength is calculated considering an intrinsic line profile shaped by the thermal broadening and the Doppler shifts based on the perturbed orbital velocity of the gas provided by the hydrodynamical simulations. To calculate the mass absorption coefficient of  $^{12}\mathrm{C}^{16}\mathrm{O}$, we adopt the transition data (transition probability $A$, lower and upper state energy $E_\mathrm{l},\,E_\mathrm{u}$, and statistical weight $g_\mathrm{u}$) provided by \citet{Goorvitch1994}.

\subsection{Characterization of CO Lines of an Eccentric Disk}

\begin{figure*}
	\centering
	\includegraphics[width=\columnwidth]{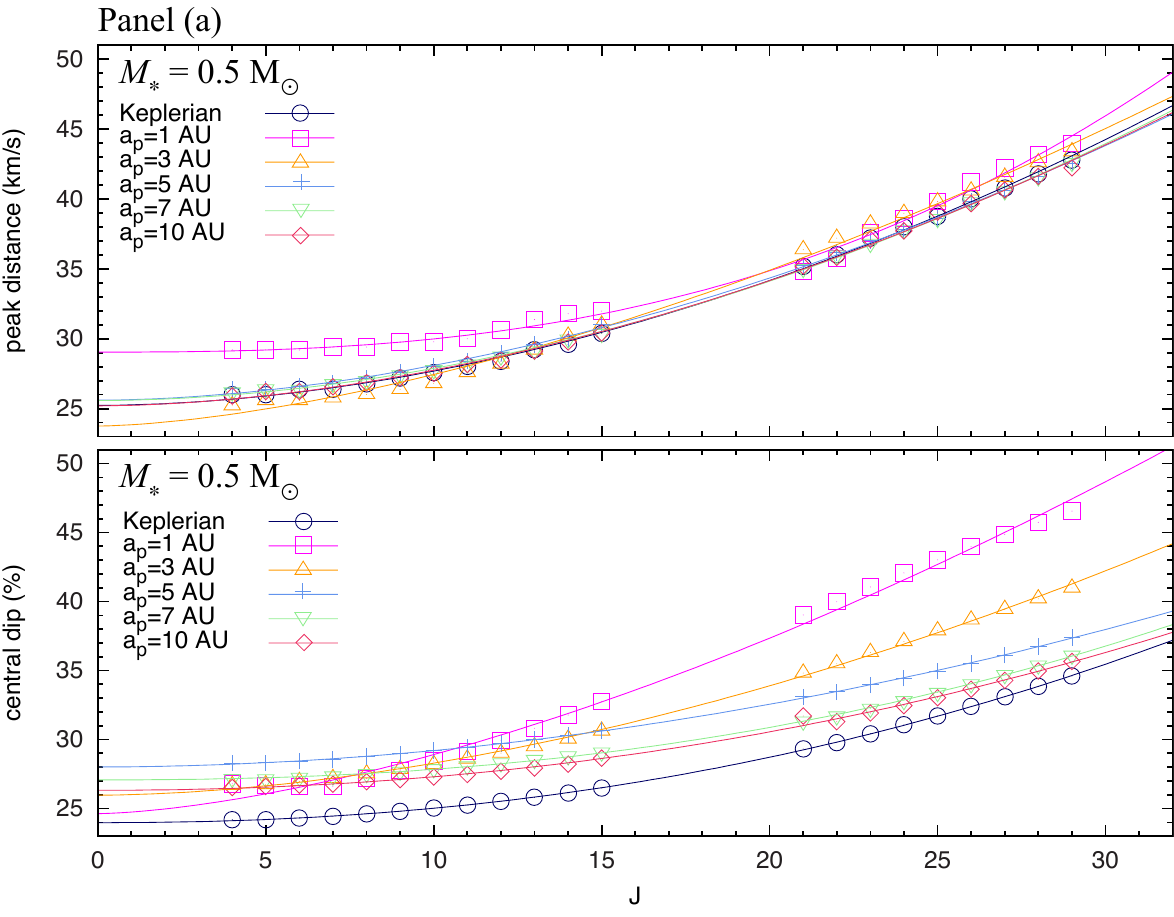}
	\includegraphics[width=\columnwidth]{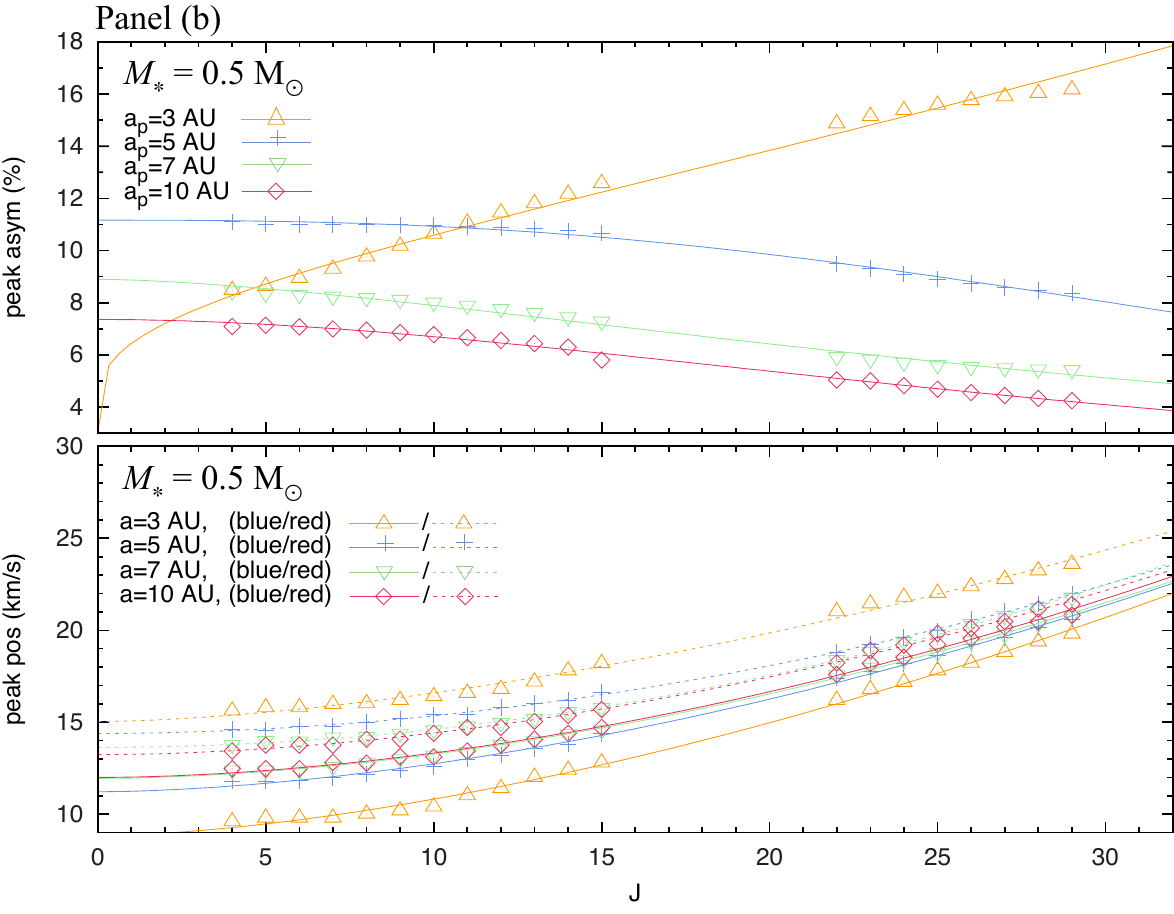}
	\includegraphics[width=\columnwidth]{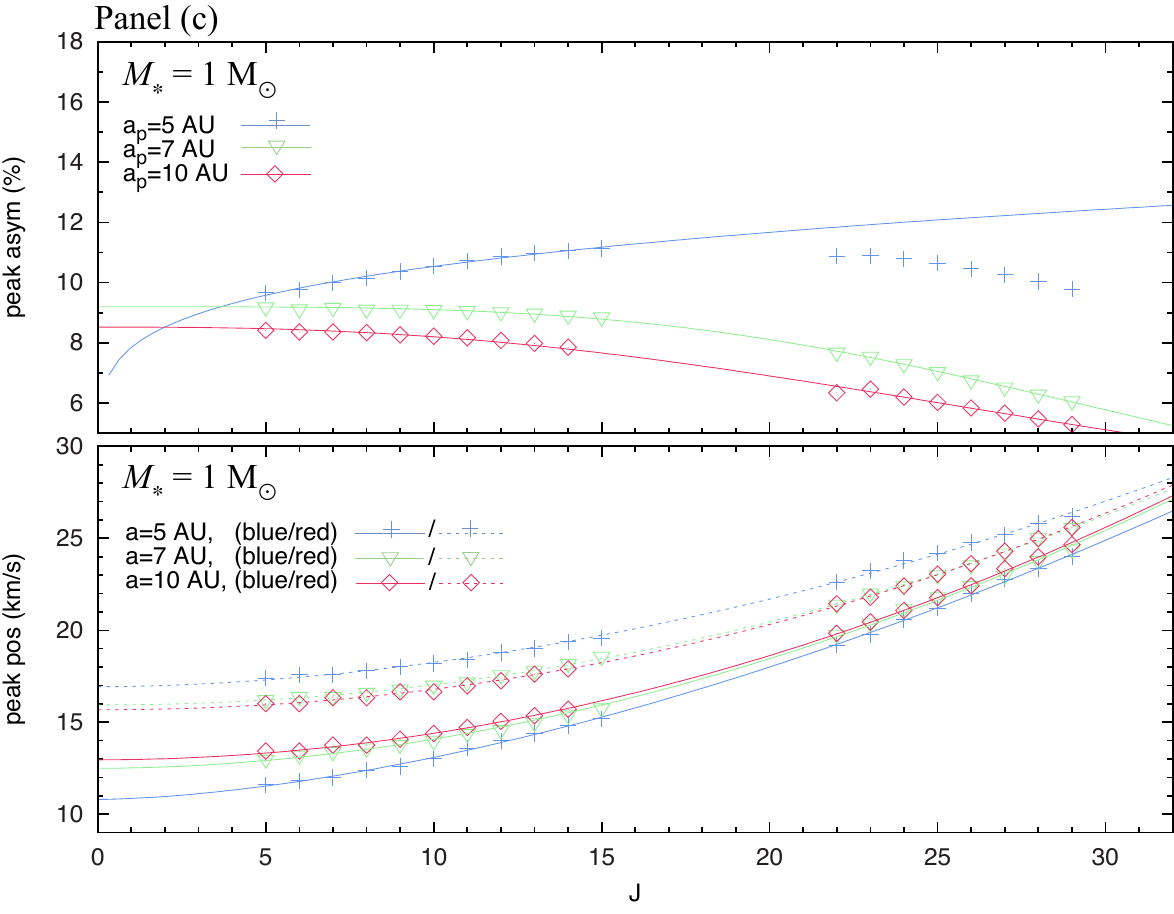}
	\includegraphics[width=\columnwidth]{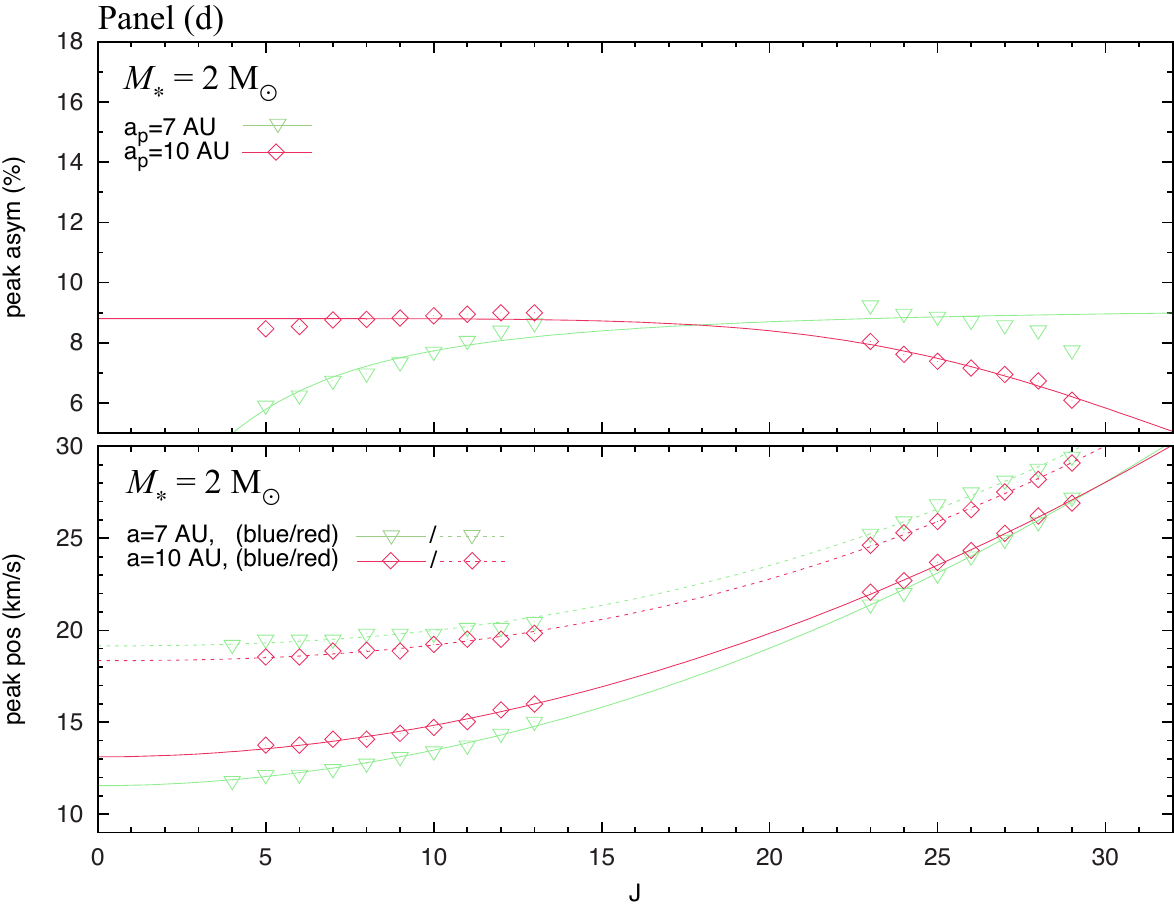}
	\caption{Line shape parameters as a function of the rotational quantum number $J$ of non-blended CO lines in the {\it P}-branch for different orbital distances of the planet. Panel (a): peak distance (upper sub-panel) and central dip (lower sub-panel)  for $0.5\,M_\odot$ central star. Panel (b)--(d): line asymmetry (upper sub-panels) and red/blue peak positions (lower sub-panels) for a $0.5,\,1$, and $2M_\odot$ central star. Only clear asymmetric lines are shown in the line asymmetry plots.}
	\label{fig:rot-diagP}
\end{figure*}

\begin{figure*}
	\centering
	\includegraphics[width=\columnwidth]{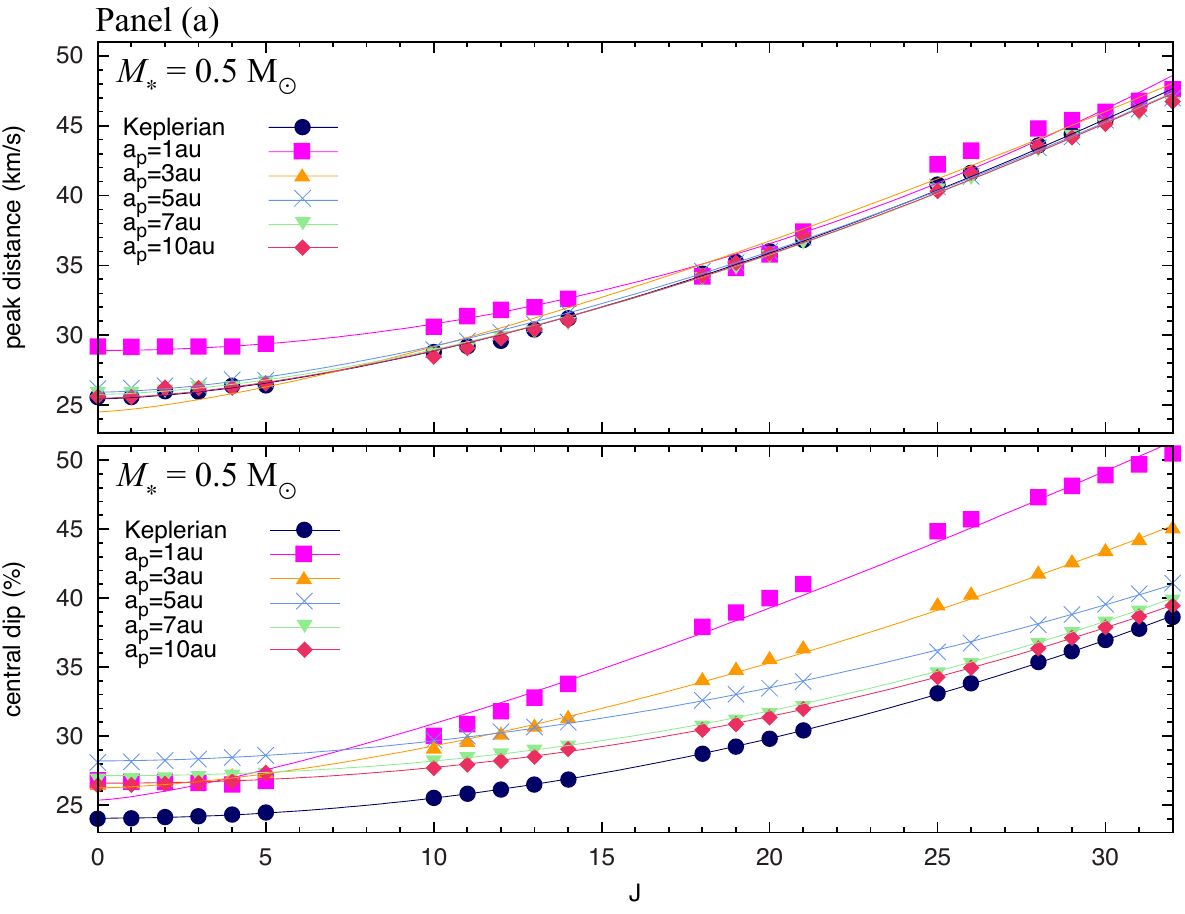}
	\includegraphics[width=\columnwidth]{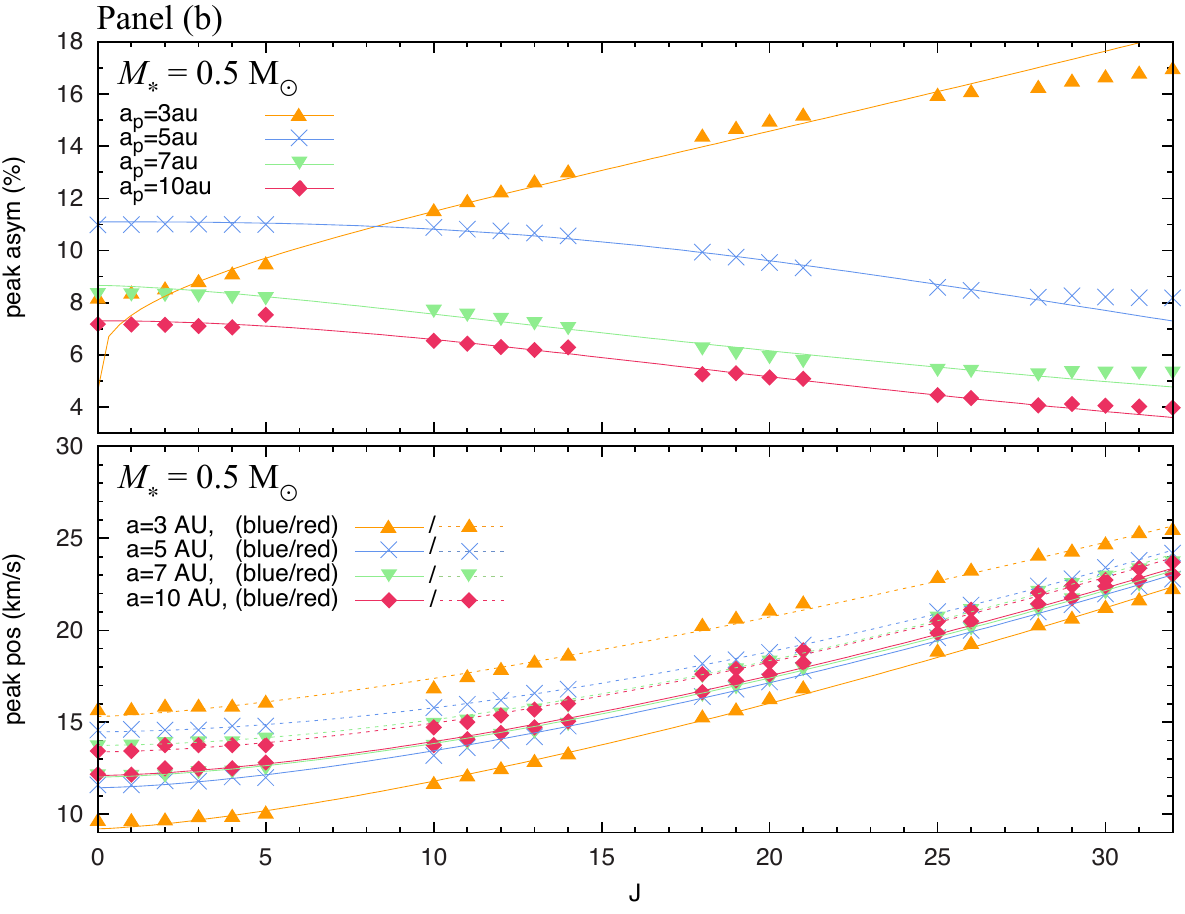}
	\includegraphics[width=\columnwidth]{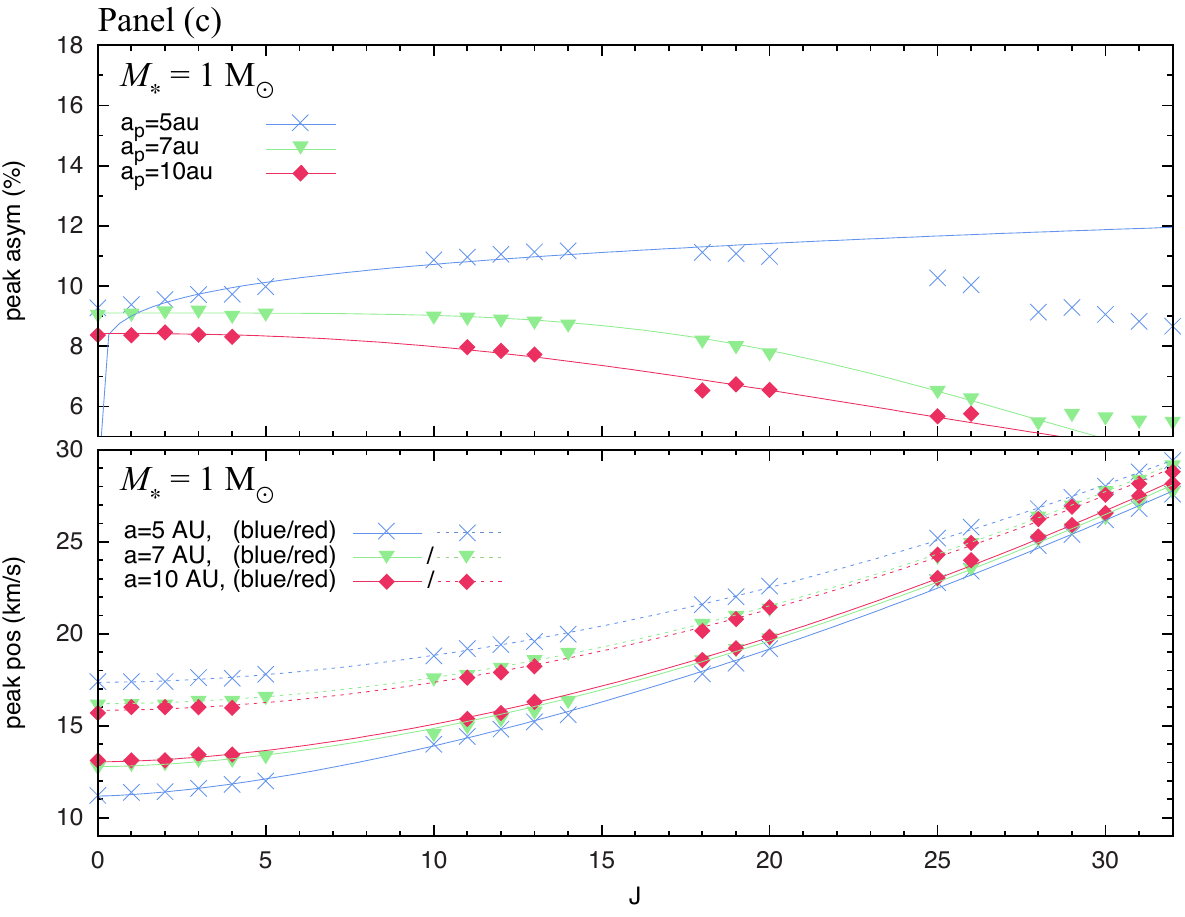}
	\includegraphics[width=\columnwidth]{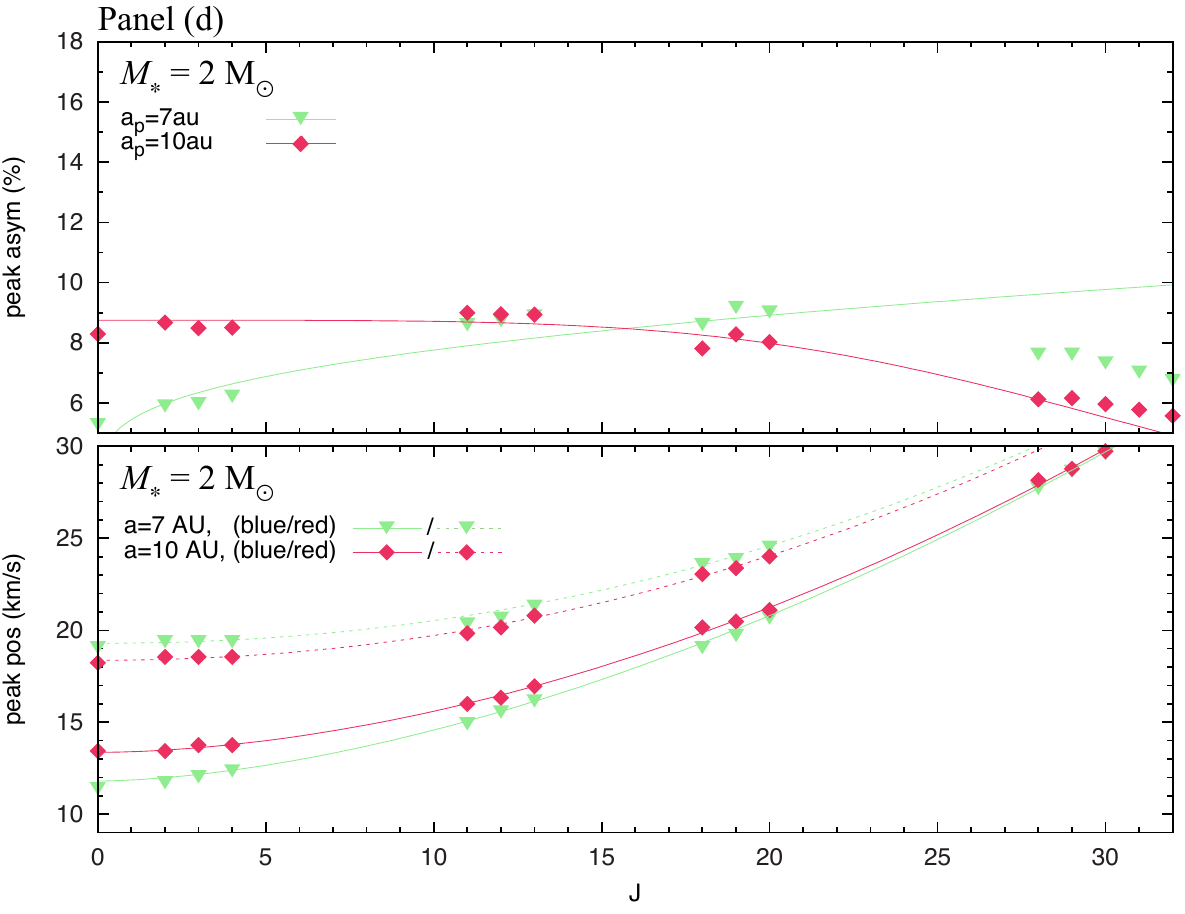}
	\caption{Same as Figure\,\ref{fig:rot-diagP}, but for {\it R}-branch CO lines.}
	\label{fig:rot-diagR}
\end{figure*}

To characterize the shape of the CO lines, we define four parameters for the continuum-normalized lines (Figure\,\ref{fig:lineprofiles}, left panel). The peak distance is $\Delta v_\mathrm{b}+\Delta v_\mathrm{r}$, where $\Delta v_\mathrm{b}$ and $\Delta v_\mathrm{r}$ are the positions of the blue and red peaks measured from the line center. The peak asymmetry is $|f_\mathrm{p(b)}-f_\mathrm{p(r)}|/f_\mathrm{max}$, where $f_\mathrm{p(b)}$ and $f_\mathrm{p(r)}$ are the line flux at the blue and red peaks and $f_\mathrm{max}$ is the maximum of the line flux. The central dip is $f_\mathrm{max}/f_0$, where $f_0$ is the line flux measured at the line center. 

We investigate the fundamental band ($V=1\rightarrow0$) of CO, as the higher order transitions ($V=2\rightarrow1\,,3\rightarrow2$, etc.) are much weaker. To avoid contamination of the eccentricity signal by high-order blending, we select transitions which have less than 1\% peak asymmetry for an unperturbed disk model. Line profiles are convolved with a $\sigma_\mathrm{res}=3\,\mathrm{km\,s^{-1}}$ Gaussian function to mimic the instrumental resolution of the Cryogenic High Resolution Echelle Spectrograph on VLT \citep{Kaeufletal2004}.

Panel (a) of Figure\,\ref{fig:lineprofiles} shows the continuum normalized (and vertically shifted by 1) profiles of some selected non-blended lines in the {\it P}-branch for an unperturbed and an eccentric model where $a_\mathrm{p}=5$\,AU and $M_*=0.5\,M_\odot$. Compared to the unperturbed Keplerian disk, the lines are asymmetric due to the supersonic disturbances in the gas velocity for eccentric disks (see detailed explanation in \citealp{Regalyetal2011}). 

Panels (b)--(d) of Figure\,\ref{fig:lineprofiles} show the strongest {\it P}(10) line\footnote{The relative strength of the CO lines depends on the temperature of the atmospheric gas and the disk interior. In our thermal model, the strongest transition in the {\it P}-branch is {\it P}(10) independent of the mass of the central star.} only for eccentric disk models with different orbital distances of the planet, and stellar masses. The lines are clearly asymmetric if $a_\mathrm{p}\geq3,\,5$, and $7$\,AU, while they have a highly distorted shape for $a_\mathrm{p}\leq1,\,3$, and $5$\,AU for an $M_*=0.5,\,1$, and $2\,M_\odot$ central star, respectively. For the highly distorted lines, the eccentricity profiles vary rapidly inside the CO thermal excitation zone (see solid curve in Figure\,\ref{fig:decprof}), which extends farther out with the stellar mass due to the higher stellar temperature.

Panel (a) of Figures\,\ref{fig:rot-diagP} and \ref{fig:rot-diagR} show the line shape parameters of non-blended $V=1\rightarrow0$ transitions in the {\it P}- and {\it R}-branch as a function of the rotational quantum number of the given transition ($J$) for both the unperturbed and eccentric disks. Both the peak distance (upper panel) and central dip (lower panel) increase with $J$. Compared to the unperturbed disk model, the peak distances are very similar (except model $a_\mathrm{p}=1\,\rm AU$); however, the central dips are significantly larger for eccentric disks. The magnitude of the central dip for a given transition firmly increases with decreasing orbital distance.

Peak asymmetries as a function of $J$ are shown in panels (a)--(c) of Figures.\,\ref{fig:rot-diagP} and \ref{fig:rot-diagR} (upper sub-panels) for $M_*=0.5,\,1$, and $2\,M_\odot$ stars, respectively. In the upper panels (peak asymmetry) only those models where the lines are clearly asymmetric are shown.\footnote{Line asymmetry cannot be defined unequivocally in models $a_\mathrm{p}\leq1,3$, and $5$\,AU for $0.5,\,1$, and $2\,M_\odot$, respectively, because the CO lines are highly distorted.} The asymmetry increases with $J$ independent of the stellar mass as long as the planet orbits closer than the outer boundary of the thermal excitation of CO ($a_\mathrm{p}\simeq 3,\,5$ and $7\, \rm AU$), while it declines with $J$ for wider planetary orbits. In the former models, the asymmetry declines if $J>20$ for 1 and $2\,M_\odot$, while it strengthens further for $0.5\,M_\odot$ stellar masses.

The excitation energy of a given transition increases with $J$, and therefore the larger $J$ is, the smaller the radial distance where that line originates. If $e_\mathrm{disk}$ has a maximum well inside the CO excitation zone, the average $e_\mathrm{disk}$ is inversely proportional to $R$ (i.e., to $J$), resulting in increasing  asymmetry with $J$. On the contrary, if $e_\mathrm{disk}$ peaks  outside the CO excitation zone, $e_\mathrm{disk}$ declines with $J$ on average, resulting in an opposite asymmetry--$J$ relation. 

Independent of the stellar mass and orbital distance, both the red and blue line peaks shift farther from the line center as $J$ increases (panels (a)--(c) of Figures\,\ref{fig:rot-diagP} and \ref{fig:rot-diagR}, lower sub-panels). The red peaks are always shifted farther from the line center than the blue peaks, meaning that the lines are off-centered. The magnitude of the off-centering increases with the stellar mass and decreasing with the orbital distance in the range of $\sim3\textrm{--}10\,\mathrm{km\,s^{-1}}$. The off-centering is caused by the excitation levels of the CO molecule (i.e., the temperature), which has the same receding and approaching velocities, being different due to the eccentric orbit of gas parcels.  Since $e_\mathrm{disk}$  grows steeper with $R$ for smaller $a_\mathrm{p}$, the difference in the level of CO excitation at the receding and approaching sides, i.e., the off-centering, is inversely proportional to $a_\mathrm{p}$. This effect is naturally magnified by the higher stellar temperature, resulting in stronger off-centering for higher mass stars.

The line asymmetry is sensitive to the disk orientation angle, thus the blue-excess changes to a symmetric, red-excess and so on, since the inner disk precesses with the $\sim360$ orbital period of the planet. This variability, however,  is hard to observe within a couple of decades as the precession period is about 250\,yr for the fastest precession observed in model $a_\mathrm{p}=1\,\rm AU$ for a $2\,M_\odot$ star. Another source of variability  might be the slight change in the disk eccentricity as the planet orbits; however, an observable signal requires a small orbital distance ($a_\mathrm{p}\leq1$\,AU; \citealp{Regalyetal2010}).

\section{Summary and Conclusions}

In this Letter, we presented fundamental band spectra of the CO molecule in circumstellar disks gravitationally perturbed  by an embedded $2.5,\,5$, and $10\, M_\mathrm{Jup}$ giant planet orbiting $M_*=0.5,\,1$, and $2\,M_\odot$ stars, respectively, at $1\,\mathrm{AU}\leq a_\mathrm{p}\leq 10\,\mathrm{AU}$. We found that the CO lines are asymmetric due to the development of disk eccentricity inside the region where CO is thermally excited.  The principal results of our study are:

\begin{enumerate}
\itemsep2pt
\parskip0pt 
\parsep0pt
\parindent0pt
\itemindent0pt

\item{A quasi-static, globally eccentric disk state develops inside the planetary orbit with $\langle {e}_\mathrm{disk}\rangle=0.2\textrm{--}0.25$ after several thousand orbits by $t\simeq40\times10^3$\,yr.}

\item{As a result, asymmetric lines are formed with magnitude of $\sim10\%\textrm{--}20\%$ measured in red versus blue peak, whose strength increases with decreasing orbital distance of the planet.}

\item{The lines are off-centered, i.e., shifted toward the higher flux peak by $\sim4\textrm{--}10\,\mathrm{km\,s^{-1}}$, such that the smaller the orbital distance of the  planet or the larger the stellar mass, the greater the shift.}

\item{The line asymmetry--$J$ slope informs us whether the giant planet orbits outside (asymmetry decreases with $J$) or inside (asymmetry increases with $J$) the thermally excited CO region. The critical orbital distance is $a_{p}\simeq3,\,5$, and $7\,\rm AU$ for an $M_*=0.5,\,1$, and $2\,M_\odot$ star, where the asymmetry--$J$ slope changes its sign.}

\item{For close planetary orbits ($a_{p}\leq1,\,3$, and $5\,\rm AU$ for an $M_*=0.5,\,1$, and $2\,M_\odot$ star), the lines are highly distorted.), the CO lines do not show clear asymmetry; rather, highly perturbed lines are formed (as was shown previously in \citealp{Regalyetal2010}) because the planet-induced gap is formed inside the CO excitation region.}

\item{The peak distances, peak positions, and magnitude of the central dip are found to be proportional to $J$ independent of the planetary and stellar masses.}
\end{enumerate}

In light of our findings (line shape dependency on $J$), we have to emphasize that care must be taken when applying line profile averaging in observational studies: averaging of only neighboring transitions (close in $J$) are suggested in which case the difference in the line shapes are  modest.  

Since the development of disk eccentricity presumably depends on radiative processes as well as the disks physical parameters (e.g., viscosity, aspect ratio, etc.) and naturally the mass of the perturbing planet, further investigation is inevitable before attempting any detailed characterization of planets from the observed line profiles.

A young ($1\textrm{--}5$\,Myr), gas-rich disk that emits asymmetric CO lines may harbor  giant planets which should be formed early in the disks life. The classical core accretion scenario predicts that the formation of giant planet takes $5\textrm{--}10$\,Myr \citep{Pollacketal1996}, which is very close to or even larger than the observed disk lifetimes \citep{Haischetal2001,Hernandezetal2007}. Recently, several theoretical studies of core accretion have reported a shorter formation time  ($1\textrm{--}5$\,Myr)  of giant planets. Planetary migration prevents the severe depletion of the feeding zone as observed in in situ calculations; however, Type\,I migration must be artificially slowed by an order of magnitude \citep{Alibertetal2005}. Assuming a protoplanetary disk with a high solid surface density of  $10\,\rm g\,cm^{-2}$ and  dust opacity in the protoplanet's envelope equals to 2\% that of interstellar material will result in a $2.5\textrm{--}3$\,Myr formation time of a Jupiter-mass planet \citep{Hubickyjetal2005,Lissaueretal2009,Movshovitzetal2010}. However, the required solid surface density is about an order of magnitude larger than that proposed by the minimum solar mass nebula model at 5.2\,AU assuming a canonical $10^{-2}$ dust-to-gas mass ratio \citep{Hayashi1981}. Note that  giant planets can be formed much faster within the context of gravitational instability theory, but it requires too high disk  masses ($>0.1\,M_\odot$) and an efficient cooling mechanism to operate \citep{Boss1997}. 

Another possible scenario could be vortex-aided planet formation in which the large-scale anticyclonic vortex developed near the disks outer dead zone edge (\citealp{Regalyetal2012}) may induce planet formation. The core of a giant planet is subject to be trapped temporarily  in the vortex vicinity \citep{Regalyetal2013}, where large amounts of dusty material are accumulated. Therefore, gas giants can be formed rapidly in both scenarios.  In the core-accretion scenario,  the slow growth regime of the giant planetary core might be circumvented by shortening the slow growth phase within the isolation ($\sim10\,M_\oplus$) and the runaway growth ($\sim20\,M_\oplus$) mass limits. In the gravitational instability scenario, the disk might be gravitationally unstable without requiring high disk mass due to the large amount of  dusty and gaseous material accumulated in the planetary trap. As the presented spectroscopic phenomenon can be applied for the characterization of planets still embedded in their host disks, it can deliver new types of empirical tests of planet formation theories.

\section*{Acknowledgments}

This project has been supported by the Lend\"ulet-2009 Young Researchers Grant of the HAS, Mag Zrt. grant No. MB08C 81013, Hungarian Grant K101393 and the City of Szombathely agreement No. K111027. Zs. R. thanks the NVIDIA Corporation for providing  TESLA C2075 cards.

\end{document}